\documentclass[11pt,twoside]{article}
\usepackage{asp2010,graphicx}

\resetcounters

\begin{document}

\title{Probing High-Column Outflows in BALQSOs Using Metastable Helium}
\author{Karen M.\ Leighly,$^1$ Adrian B. Lucy,$^1$ Matthias
  Dietrich,$^2$ Donald Terndrup,$^{2,3}$ and Sarah C. Gallagher$^4$
\affil{$^1$Homer L.\ Dodge Department of Physics and Astronomy, The
  University of Oklahoma, 440 W.\ Brooks St., Norman, OK 73019,
  $^2$Department of Astronomy, The Ohio State University, 4055 
  McPherson Lab, 140 W.\ 18th Ave, Columbus, OH 43210, $^3$National
  Science Foundation, 4201 Wilson Blvd., Arlington, VA 22230,
  $^4$Department of Physics and Astronomy, The University of  
  Western Ontario, 1151 Richmond Street, London, ON N6A 3K7, Canada}} 

\begin{abstract}
Outflows are believed to be ubiquitous and fundamentally important in
active galaxies. Despite their importance, key physical properties of
outflows remain poorly unconstrained; this severely limits study of the
acceleration process. It is especially difficult to
constrain the column density since most of the lines are
saturated. However, column densities can be measured using ions that
are expected to be relatively rare in the gas, since they are least
likely to be saturated. Phosphorus, specifically the P\,{\sc v}
doublet at 1118 and 1128\AA\/, is generally regarded as a useful
probe of high column densities because of its low abundance. We have
found that the metastable neutral helium triplet is an equally
valuable probe of high column densities in BALQSOs. The significant
advantage is that it can be observed in  the infrared (He\,{\sc
  i}*$\lambda 10830$) and the optical (He\,{\sc   i}*$\lambda 3888$)
bands from the ground in low-redshift ($z<1.2$) objects. 

We report the discovery of the first He~{\sc i}*$\lambda 10830$ BALQSO
FBQS J1151+3822, and discuss constraints on the column density
obtained from the optical and IR He~{\sc i}* lines.  In addition, a
new observation revealing  Mg\,{\sc ii} and Fe\,{\sc   ii} absorption
provides further constraints, and {\it   Cloudy} modeling of 
He~{\sc i}*, Mg\,{\sc ii} and Fe\,{\sc   ii} suggests that the
difference between LoBALs and FeLoBALs is column density along the
line of sight.  

\end{abstract}

\section{The Remarkable Properties of He\,{\sc i}*}
Neutral helium has interesting atomic structure that makes it a
useful astrophysical tool.  The 2s level of the triplet state is
metastable with a lifetime of 2.2 hours, and it acts as a 
second ground state.   The 2s level lies 19.75~eV above the ground
state, so it is populated by recombination.  It is depopulated
predominantly by collisions in photoionized gas.  This means that He~{\sc i}*
absorption measures the He$^+$ column, and He\,{\sc i}* absorption occurs
along with other high-ionization absorption lines including C~{\sc iv},
Si~{\sc iv}, and P~{\sc v}.

\begin{figure}[t]
\epsscale{1.0}
\plotone{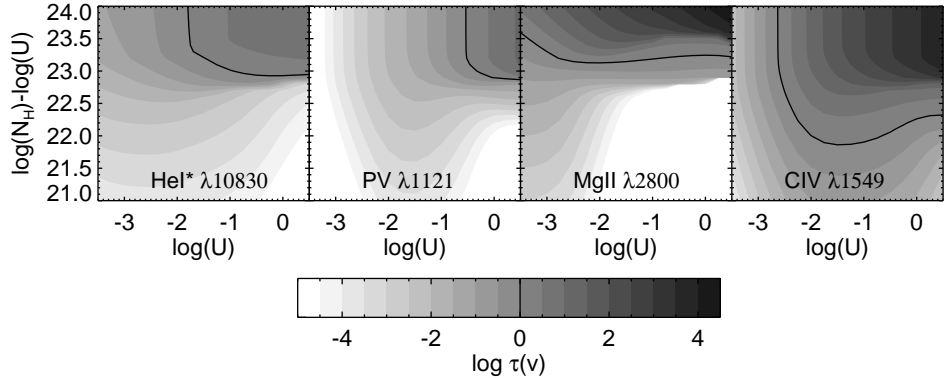}
\caption[]{Predicted opacity contours for a square absorption
  line with a width of 10,000$\rm \, km\, s^{-1}$.  He~{\sc
    i}*$\lambda 10830$ has nearly the same opacity as P~{\sc v} over a
  broad area of parameter space.  In contrast, at high values of
  $\log (N_H)-\log(U)$, lines from common ions such as C~{\sc
    iv}$\lambda 1549$ would be saturated. \label{comparison}}
\end{figure}

It has been known for $\sim$15 years that many quasar absorption lines
are saturated but not black because the absorber only partially covers
the continuum source \citep[e.g.,][]{hamann98}.  So, to estimate
column densities, one needs absorption lines that are not easily
saturated, i.e., from rare ions.  Then, two or more transitions from
the same lower level have ratios fixed by atomic physics, and
measurements of optical depths of lines from such transitions can be
used to solve for the covering fraction.  Phosphorus is a
low-abundance element (765 times less abundant than carbon).  The
P\,{\sc v} resonance doublet at 1118 and 1128\AA\/ is a good probe of
high-column outflows \citep{hamann98}.

Although helium is abundant in astronomical 
gas, He~{\sc i}* is rare.  One of every $\sim$175,000 He$^+$
atoms is He~{\sc i}* in a typical plasma \citep{clegg87}.  
The He~{\sc i}*$\lambda$10830 (2s$\rightarrow$2p) and
He~{\sc i}*$\lambda$3889 (2s$\rightarrow$3p) lines are suitable for
measuring  
the He$^+$ covering fraction and column density.  As shown in
Fig.~\ref{comparison}, He~{\sc i}* has very similar opacity as P~{\sc
  v}.  But He~{\sc i}* has the added advantage that it is observable
from the ground in the optical and IR in low redshift objects.  

He~{\sc i}* has another important advantage over other lines.  A line's
opacity is proportional to $\lambda f_{ik}$, where $f_{ik}$ is the
oscillator strength.  So the optical depth ratio of two transitions
from the same level is equal to the $\lambda f_{ik}$ ratio.  This
ratio is $\sim$2 for hydrogen-like resonance doublets (including
P~{\sc v}).  In contrast, the ratio of $\lambda f_{ik}$ for
He~{\sc i}*$\lambda$10830 to He~{\sc i}*$\lambda$3889 is 23.3.  This
means that these two lines provide a large dynamic range
for measuring the spatially-averaged optical depth.  Specifically,
He~{\sc i}*$\lambda 10830$ is sensitive to lower column densities, and
He~{\sc i}*$\lambda 3889$ takes over at higher column densities.
Simulations show that the average column density can be accurately
recovered over almost 2.5 orders of magnitude \citep[][Fig.\ 19]{leighly11}.  

\section{FBQS~J1151$+$3822: The First He\,{\sc i}*$\lambda$10830
  BALQSO} 

We observed FBQS~J1151$+$3822 (z=0.3344, $m_v=15.7$, $M_V=-25.6$) in
March 2008 using SpeX on the NASA Infrared Telescope Facility (IRTF)
as part of another project.  We serendipitously discovered a prominent
broad absorption feature that could only be attributed to
He~{\sc i}*$\lambda 10830$.  While absorption from He~{\sc i}*$\lambda
3889$ has been observed in other objects, this was the first report of
a broad absorption line in He~{\sc i}*$\lambda 10830$.  This work is
described in \citet{leighly11}. 

\begin{figure}[t]
\epsscale{1.0}
\plotone{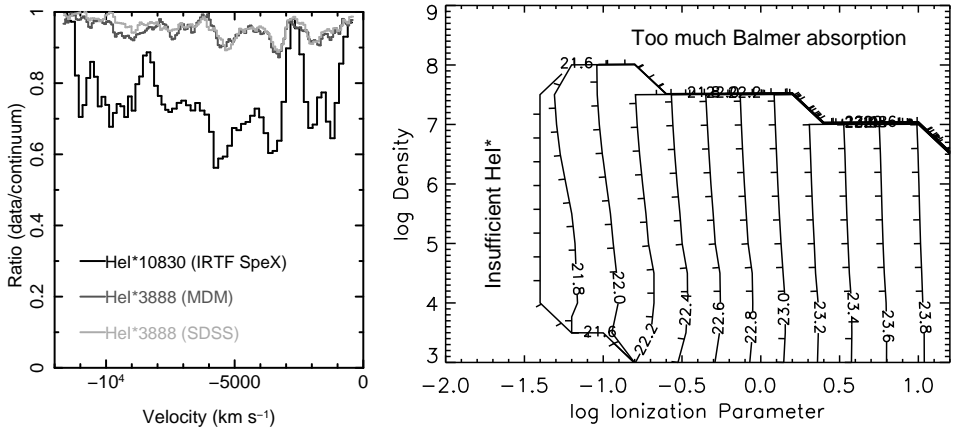}
\caption[]{{\it Left:}  The ratio of data to continuum as a function
  of velocity for the He~{\sc i}* lines in FBQS~J1151$+$3822.   {\it
    Right:} Contours of the total log hydrogen column  density [$\rm
    cm^{-2}$] required   to produce the measured average He~{\sc i}* 
  plotted as a function of the log of the ionization parameter $U$ and
  the log of the hydrogen density [$\rm cm^{-3}$].    See \citet{leighly11} 
  for details. \label{hei_for_fbqs}} 
\end{figure}

Absorption from He~{\sc i}* at 3889\AA\/ is not obvious in
FBQS~J1151$+$3822 because of strong Fe~{\sc ii} emission.  However,
the 3889\AA\/ component could be recovered using an Fe~{\sc ii}
template model.  The ratio of the data to continuum is shown for the
He~{\sc i}*$\lambda 10830$ and  He~{\sc i}*$\lambda 3889$ lines in
Fig.~\ref{hei_for_fbqs}.  We used standard partial covering analysis
\citep[e.g.,][]{hamann97,arav05} to derive the optical depth and
covering fraction as a function of velocity. Integration over the
optical depth yields a log He~{\sc i}* column density of $\sim$15.6
[$\rm cm^{-2}$].   The spatially-averaged log column density
\citep[i.e., $C_f \tau$, where   $C_f$ is the covering fraction,
  e.g.,][]{arav05} is $\sim$14.9. There are no Balmer  absorption
lines in the spectrum, so we obtain an upper limit on hydrogen in the
$n=2$ level.     

{\it Cloudy} modeling was  used to determine the total hydrogen column
density, where solar abundances were assumed throughout.  We show in
Fig.~\ref{hei_for_fbqs} the contours of total 
hydrogen column density required to produce the observed average
He~{\sc i}* column density.  Low ionization parameters produce
insufficient He~{\sc i}* because the He$^+$ region is too thin.
High densities violate the Balmer absorption upper limit.  The column
densities estimated for FBQS~J1151$+$3822 are comparable to or larger
than those from several objects presented by \citet{dunn10} (also
obtained using partial covering analysis).  This demonstrates the
sensitivity of He~{\sc   i}* to high column densities.    

\section{Mg~{\sc ii} and Fe~{\sc ii} Absorption in FBQS J1151$+$3822}

\begin{figure}[t]
\epsscale{1.0}
\plotone{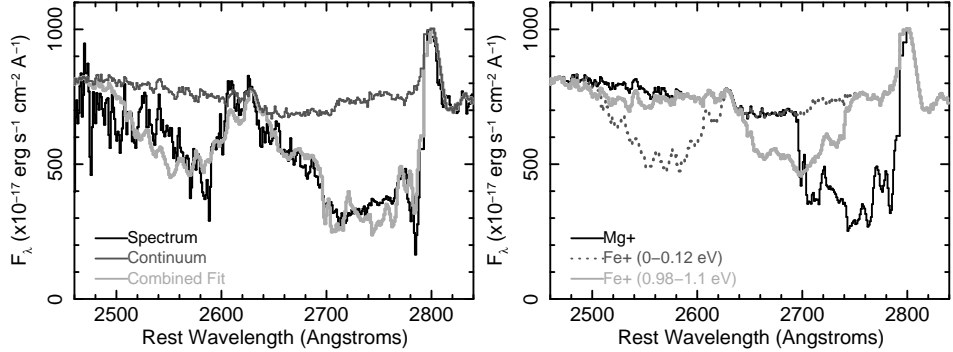}
\caption[]{{\it Left:} The KPNO 4m spectrum of FBQS~J1151$+$3822,
  showing the best fit model for a  preliminary fit   to the Mg~{\sc
    ii}/Fe~{\sc ii}   feature using a   model created with the He~{\sc
    i}*$\lambda 10830$   absorption profile.  {\it Right:} The three
separate  absorption components in the model  (Lucy et al.\ in
prep.). \label{mgii}} 
\end{figure}

We observed FBQS~J1151$+$3822 in May 2011 at the KPNO 4-meter telescope
using the RC spectrograph.  We detected, for the first time, a broad
absorption feature shortward of the Mg~{\sc ii} emission line (Fig.~\ref{mgii}; Lucy et
al. in prep.).  

Clearly, Mg~{\sc ii} absorption is present in the spectrum.  We conclude that
absorption from Fe~{\sc ii} and excited state Fe~{\sc ii} is also
present.  We construct absorption profiles for Mg~{\sc ii} and Fe~{\sc
  ii} using the He~{\sc i}*$\lambda$10830  profile, thereby assuming
that all of these lines are produced in the same gas (the
justification for this assumption will be discussed in Lucy et al.\ in
prep.).  The Fe~{\sc ii} divides
nicely into ground state and low ionization 
lines ($\lambda < 2600$\AA\/), and higher ionization lines
(2600--2750\AA\/).  We fit the spectrum with these profiles
(Fig.~\ref{mgii}). Integrating over the resulting apparent optical
depths yields apparent log column densities 
of 15.0, 14.6 and 13.7 [$\rm cm^{-2}$] for Mg~{\sc ii}, ground and
low-excitation Fe~{\sc ii}, and high excitation Fe~{\sc ii},
respectively.  For  shallow features like the Fe~{\sc ii} features,
\citet{leighly11} show that the spatially-averaged column is
approximately  equal to the column obtained from the apparent optical
depth.   

\begin{figure}[t]
\epsscale{1.0}
\plotone{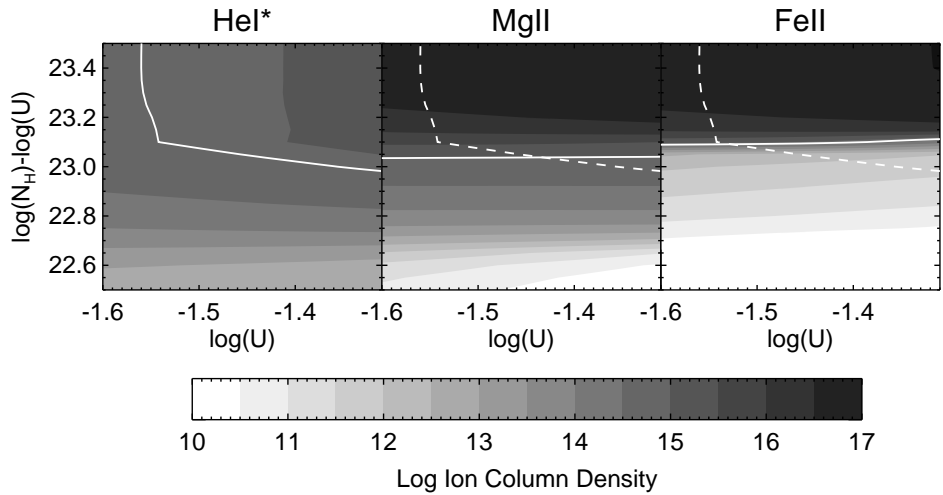}
\caption[]{Column densities obtained using {\it Cloudy}.  The solid
  white line shows the location in parameter space where the column
  density matches the measured value.  The dashed line shows the
  He~{\sc i}* average column on the other plots to guide the
  eye.  While the He~{\sc i}* and Mg~{\sc ii} values are consistent
  for $\log U \approx -1.4$, Fe~{\sc ii} requires $\log U = -1.55$,
  indicating that Mg~{\sc ii} is saturated.  \label{cloudy}} 
\end{figure}

Our measurements of Mg~{\sc ii} and Fe~{\sc ii} place new strong
constraints on the ionization parameter and therefore the column
density. Fig.~\ref{cloudy} shows {\it Cloudy} simulations of column
densities of He~{\sc i}*, Mg$^+$ and Fe$^+$.  We find that sufficient 
Mg~{\sc ii} is produced for $\log(U) \leq -1.4$. However, sufficient
Fe$^+$ is attained only at $\log(U)\approx -1.55$, where He~{\sc i}* is ionization
bounded (i.e., the  He~{\sc i}* predicted in a semi-infinite slab is
equal to the measured value).   This is a consequence of the atomic
properties of Mg and Fe. In the H~{\sc ii} region, Mg$^+$ and Fe$^+$ are
produced by recombination.   The +2$\rightarrow$+3 ionization
potential for Mg is high, 80.1~eV, which means that there is plenty of
Mg$^{+2}$, and therefore Mg$^{+1}$ in the H~{\sc ii} region, more than
enough to create the observed broad absorption line.  But the similar
ionization   potential for Fe is 30.7~eV, implying that iron is
distributed among many ionization states in the H~{\sc ii} region.
This means that a 
much higher column is needed to observe Fe~{\sc ii}  compared with
Mg~{\sc ii}. Therefore, the difference between LoBALs (objects with
only Mg~{\sc ii}) and FeLoBALs (objects that also have Fe~{\sc ii}) 
may be simply a difference in column density along the line of sight.  

The strongly-constrained ionization parameter provides strong
constraints for the column density as well.  For example, for log gas
density  $\log n=6.5$ [$\rm cm^{-3}$], {\it Cloudy} models for He~{\sc
  i}* and Fe~{\sc ii} match the measured values of those components
where  $\log(N_H)=21.65$ [$\rm cm^{-2}$].  Mg~{\sc ii} far exceeds the
measured value at this column density, implying that the Mg~{\sc ii}
line is saturated.    

The value of $\log (n)$ is constrained to be between 5.0 and 7.5 
by the  lack of Balmer absorption and the presence of excited state
Fe~{\sc ii}.  If we assume that the outflow is accelerated by radiative line driving,
\citet{leighly11} show that $\log(n)>7.0$.  This constrains
$\log(N_H)=21.7$ [$\rm cm^{-2}$], the absorption-line gas radius of
12--21~pc, mass flux between 18 and 31 solar masses/yr, log kinetic
luminosity between 44.2 and 44.45 [$\rm erg\, s^{-1}$], and  ratio of
kinetic to bolometric luminosity to between 0.3 and 0.5\%.
\citet{he10} postulate that a ratio of kinetic to bolometric
luminosity as low as 0.5\% may be sufficient to truncate star
formation in AGN host galaxies.  In that case, the outflow displayed
in FBQS 1151+3822 may be just sufficient.   

\section{Summary and the Future}

He~{\sc i}* has been shown to be a powerful probe of
high-column-density outflows  in low redshift quasars.   High column
 outflows are interesting because they may carry the most
kinetic energy, and they potentially provide the greatest challenges to
acceleration models.  Low redshift objects are interesting since they
provide potentially valuable targets for imaging.  Furthermore,
 He~{\sc i}* lines can be observed from the ground, which means that
 uniform samples can be  identified and  studied.  

We have several followup projects in progress.  Using IRTF, we 
observed a small sample of low-redshift quasars known to be BALQSOs.
Some, like Mrk~231, have He~{\sc   i}*$\lambda 10830$ absorption, and
others do not.   We identified 18 low-redshift quasars that have identifiable
He~{\sc i}*$\lambda 3889$ in their SDSS spectra, and we have observed
them using LBT Luci and/or Gemini GNIRS.  This sample will allow a
systematic investigation of covering fraction and column density as a
function of other outflow parameters. 

\acknowledgements KML, ABL support: NSF AST-0707703, MD: AST-0604066.


\begin{thebibliography}{}
\expandafter\ifx\csname natexlab\endcsname\relax\def\natexlab#1{#1}\fi
\expandafter\ifx\csname url\endcsname\relax
  \def\url#1{\texttt{#1}}\fi
\expandafter\ifx\csname urlprefix\endcsname\relax\def\urlprefix{URL }\fi
\providecommand{\eprint}[2][]{\url{#2}}

\bibitem[{{Arav} et~al.(2005){Arav}, {Kaastra}, {Kriss}, {Korista}, {Gabel}, \&
  {Proga}}]{arav05}
{Arav}, N., {Kaastra}, J., {Kriss}, G.~A., {Korista}, K.~T., {Gabel}, J., \&
  {Proga}, D. 2005, \apj, 620, 665

\bibitem[{{Clegg}(1987)}]{clegg87}
{Clegg}, R.~E.~S. 1987, \mnras, 229, 31P

\bibitem[{{Dunn} et~al.(2010){Dunn}, {Bautista}, {Arav}, {Moe}, {Korista},
  {Costantini}, {Benn}, {Ellison}, \& {Edmonds}}]{dunn10}
{Dunn}, J.~P., {Bautista}, M., {Arav}, N., {Moe}, M., {Korista}, K.,
  {Costantini}, E., {Benn}, C., {Ellison}, S., \& {Edmonds}, D. 2010, \apj,
  709, 611

\bibitem[{{Hamann}(1998)}]{hamann98}
{Hamann}, F. 1998, \apj, 500, 798

\bibitem[{{Hamann} et~al.(1997){Hamann}, {Barlow}, {Junkkarinen}, \&
  {Burbidge}}]{hamann97}
{Hamann}, F., {Barlow}, T.~A., {Junkkarinen}, V., \& {Burbidge}, E.~M. 1997,
  \apj, 478, 80

\bibitem[{{Hopkins} \& {Elvis}(2010)}]{he10}
{Hopkins}, P.~F., \& {Elvis}, M. 2010, \mnras, 401, 7

\bibitem[{{Leighly} et~al.(2011){Leighly}, {Dietrich}, \& {Barber}}]{leighly11}
{Leighly}, K.~M., {Dietrich}, M., \& {Barber}, S. 2011, \apj, 728, 94

\end{thebibliography}
\end{document}